# Frame dragging with optical vortices

J. Strohaber

Texas A&M University, Department of Physics, College Station, TX 77843-4242, USA

*Corresponding author: jstroha1@physics.tamu.edu

General Relativistic calculations in the linear regime have been made for electromagnetic beams of radiation known as optical vortices. These exotic beams of light carry a physical quantity known as optical orbital angular momentum. It is found that when a massive spinning neutral particle is placed along the optical axis, a phenomenon known as inertial frame dragging occurs. Our results are compared with those found previously for a ring laser and an order of magnitude estimate of the laser intensity needed for a precession frequency of 1 Hz is given for these "steady" beams of light.

In pre-relativity, Newton's law of universal gravitation was used to quantify the gravitational force between massive point objects. From this elementary theory of gravity, it is well known that mass is the source of gravitational fields, and those gravitational fields act only upon massive particles [1,2]. In contrast, in 1804 the German physicist Soldner used Newton's corpuscular (little particle) theory of light to show that starlight skimming the sun could be deflected about a straight path. This result is often referred to as the "Newtonian prediction" [2,3]. It was not until the birth of General Relativity (GR) that Einstein theoretically demonstrated that the trajectories of photons could be influenced by a gravitational field and in the case of the deflection of starlight by the sun he gave the correct result which is twice the Newtonian prediction [4]. In May of 1919, this deflection was observed by Eddington during a total eclipse [5]. More recently, the deflection of light around massive objects has been exploited to detect and investigate astronomical phenomenon by a GR effect known as gravitational lensing [6].

Even more fascinating, in GR, the geometry of spacetime depends on the energy-momentum configure of a source, and since light carries energy and momentum, light also acts as a source of gravity. In 1931, Tolman showed that "thin pencils of light" produce a gravitational influence on test rays and particles [7]. Scully later extended upon the earlier work of Tolman *et. al.,* by calculating the gravitational coupling between subluminal laser pulses [8]. More recently, Mallett generalized upon these earlier works by solving Einstein's equations in the weak-field limit for a ring laser configuration [9]. These calculations demonstrated a phenomenon known as inertial frame dragging. Of further interest, in a later controversial publication, Mallet found exact solutions of the Einstein field equations and showed that the exterior metric contained closed timelike curves (CTCs) [10,11]. These light-induced CTCs are conjectured by Mallett to be the foundation for a time machine. In this letter, we extend upon the work of Mallett by calculating the weak gravitational influence from a Laguerre-Gaussian ($LG_p^\ell$) laser beam on a massive spinning neutral particle. These exotic beams of light known as optical vortices (OVs) have generated considerable interest in the scientific community because they carry a physical quantity known as optical orbital angular momentum (OAM) [12]. It is the goal of this paper to calculate the coupling of this unique form of radiation with a spinning test particle and to compare the results with those found in the case of a ring laser.

Under the Lorentz gauge condition $\partial_\alpha A^\alpha = 0$ and for source-free vacuum electromagnetic radiation, the four-potential satisfies the homogeneous wave equation $\partial^\mu \partial_\mu A^\alpha = 0$. The amplitude of paraxial beams are solutions of the scalar paraxial wave equation (PWE), which is ultimately derived from the wave equation. To find the electromagnetic fields of paraxial beams, the scalar and vector potentials are chosen as $A^t = A_0 \psi$ and $\vec{A} = A_0 \psi (\alpha \hat{e}_x + \beta \hat{e}_y)$ respectively [13]. Here $\alpha$ and $\beta$ are polarization parameters such that $\alpha = 1$, $\beta = 0$ ($\alpha = 0$, $\beta = 1$) correspond to polarization along the x-direction (y-direction), and $\psi$ is a solution to the PWE. Using Maxwell's equations and the Lorentz gauge condition, the electric and magnetic fields within the paraxial approximation are,

$$\vec{E} = E_0 \left[ i\left(\alpha \hat{e}_x + \beta \hat{e}_y\right)\psi - \frac{1}{k}\left(\alpha \frac{\partial \psi}{\partial x} + \beta \frac{\partial \psi}{\partial y}\right)\hat{e}_z \right] e^{i(kz-\omega t)}$$
$$\vec{B} = B_0 \left[ -i\left(\beta \hat{e}_x - \alpha \hat{e}_y\right)\psi + \frac{1}{k}\left(\beta \frac{\partial \psi}{\partial x} - \alpha \frac{\partial \psi}{\partial y}\right)\hat{e}_z \right] e^{i(kz-\omega t)} \quad . \quad (1)$$

Three families of solutions $\psi$ have been found by separation of variable. In Cartesian coordinates the solutions are the Hermite-Gaussian modes, and in cylindrical polar and elliptical coordinates they are the Laguerre- and Ince-Gaussian modes respectively. For the present work, the radiation field of the $\text{LG}_p^\ell$ beams will be of interest as the source of the gravitational influence,

$$\psi = \frac{w_0}{w}\left(\frac{\sqrt{2}r}{w}\right)^{|\ell|} L_p^{|\ell|}\left(\frac{2r^2}{w^2}\right)\exp\left(-\frac{r^2}{w^2}\right)\exp\left(-i\frac{kr^2}{2R}\right)e^{-i\ell\theta}e^{-i(2p+|\ell|+1)\Psi_G} \quad . \quad (2)$$

Here $w(z) = w_0\sqrt{1 - z^2/z_0^2}$ is the beam size, where $w_0$ and $z_0 kw_0^2/2$ are the waist and Rayleigh range respectively. $R(z) = z + z_0^2/z$ is the radius of curvature, $\Psi_G(z) = \arctan(z/z_0)$ is the Gouy phase and $L_p^{|\ell|}$ are the associated Laguerre polynomials.

Using Eqs. (1) and (2) and a change of basis, it is straightforward to calculate the Poynting vector $\vec{S} = (\vec{E} \times \vec{B})/\mu_0$ for the $\text{LG}_p^\ell$ beams in cylindrical polar coordinates [12],

$$\vec{S} = |E_0|^2 \frac{1}{c\mu_0}\left[\frac{r}{R(z)}\hat{e}_r + \frac{1}{k}\left(\frac{\ell}{r} - \sigma_z \frac{1}{2}\frac{\partial}{\partial r}\right)\hat{e}_\theta + \hat{e}_z\right]|\psi|^2 \quad . \quad (3)$$

Here the photon helicity parameter is $\sigma_z = i(\alpha\beta^* - \alpha^*\beta)$, where $\sigma_z = \pm 1$ is for right and left circularly polarized light and $\sigma_z = 0$ is for linearly polarized light. In addition to the longitudinal component of energy flow normally encountered with plane waves, paraxial beams in general have radial and azimuthal components. Integral curves of the Poynting vector "spiral" around the optical axis of the beam—suggesting that these curves are the paths traveled by the photons. The radial component is inversely proportional to $R(z)$, which is an inescapable consequence of the diffraction of finite-sized beams. For OVs, the azimuthal component is dependent upon the angular momentum mode number $\ell$ in addition to the helicity parameter $\sigma_z$. In general, the azimuthal component in Eq. (3) is small compared to the longitudinal component; however, we will demonstrate that the magnitude presents no difficulties relative to results found in Ref [9], where the longitudinal components were the sole contributor to the Poynting vectors.

For weak gravitational fields, the full nonlinear Einstein equations can be linearized as $\partial_\lambda \partial^\lambda h_{\mu\nu} = -\kappa\left(T_{\mu\nu} - \eta_{\mu\nu} T/2\right)$, where $T^{\mu\nu} = -\left(F^{\mu\alpha} F^\nu_\alpha - \eta^{\mu\nu} F_{\alpha\beta} F^{\alpha\beta}/4\right)/\mu_0$ is the energy-momentum tensor and $F_{\mu\nu} = \partial_\mu A_\nu - \partial_\nu A_\mu$ is electromagnetic tensor. It will be seen that the trace of the energy-momentum tensor is zero $T = \eta^{\alpha\beta} T_{\alpha\beta} = 0$, thereby reducing the linearized field equation to $\partial_\lambda \partial^\lambda h_{\mu\nu} = -\kappa T_{\mu\nu}$ [8,9]. As a note, in the linear approximation, Einstein's equations are analogous to potential problems typically encountered in graduate textbooks on electrodynamics [13, 14]. The solutions for the metric perturbation in the linearized version of the Einstein field equations can be approximated as [7—9],

$$h_{\mu\nu}(x) = -\kappa \int d^3 x' G(x,x') T_{\mu\nu}\left(x', t - |x - x'|\right). \tag{4}$$

Here $\kappa = 8\pi G/c^4$ (where the gravitational constant is $G = 6.67 \times 10^{-11}$ m$^3$/kg/s$^2$), $G(x,x') = 1/\left(4\pi |x - x'|\right)$ is the Green's function and $T_{\mu\nu}$ is the energy-momentum tensor. Using results from Eqs. (1) and (3) and noticing that $E_x = cB_y$ and $E_y = -cB_x$, the energy-momentum tensor for $\text{LG}_p^\ell$ beams with angular and radial "quantum" numbers $\ell$ and $p$ is found in the paraxial approximation to be,

$$T^{\mu\nu} = \frac{1}{c}\begin{bmatrix} S_z & S_x & S_y & S_z \\ S_x & 0 & 0 & S_x \\ S_y & 0 & 0 & S_y \\ S_z & S_x & S_y & S_z \end{bmatrix}. \tag{5}$$

The energy-momentum tensor in Eq. (5) is symmetric, and conveniently within the paraxial approximation, its elements can be written in terms of the Cartesian components of the Poynting vector in Eq. (3).

To test for the effects of inertial frame dragging, a spinning test particle will be placed along the optical axis of the beam [Fig. 1]. With this in mind, it is computationally advantageous to inspect the form of the General Relativistic spin equations before evaluating the integrals in Eq. (4). The General Relativistic spin equations in the slow motion and weak field approximates are given by [9,15]

$$\frac{ds_i}{dt} = c\Gamma_{i0}^k s_k - \Gamma_{i0}^0 s_j v^j + \Gamma_{il}^k s_k v^l - c^{-1}\Gamma_{ik}^0 s_j v^k v^j, \tag{6}$$

where $\Gamma_{\alpha\beta}^\gamma$ are the Christoffel connection coefficients, and $s_i$ and $v_i$ are the spin and velocity vectors respectively. For a stationary particle, Eq. (6) reduces to $ds_i/dt = c\Gamma_{i0}^k s_k$ and the required Christoffel symbols are given by $2\Gamma_{i0}^k = \eta^{kj}\left(h_{ji,0} + h_{j0,i} - h_{i0,j}\right)$. In calculating the energy-momentum tensor given in Eq. (5), a continuous-wave field was time averaged, and because the integral in Eq. (4) will be calculated for a "steady" beam of light between $z = -L/2$ and $z = L/2$, the connection coefficients take the simplified form $2\Gamma_{i0}^k = \eta^{kj}\left(h_{j0,i} - h_{i0,j}\right)$. Using these conditions, the spatial components of Eq. (6) can be written as,

$$\frac{2}{c}\frac{ds_x}{dt} = \eta^{yy}\left(h_{0y,x} - h_{x0,y}\right)s_y + \eta^{zz}\left(h_{0z,x} - h_{x0,z}\right)s_z$$
$$\frac{2}{c}\frac{ds_y}{dt} = \eta^{xx}\left(h_{0x,y} - h_{y0,x}\right)s_x + \eta^{zz}\left(h_{0z,y} - h_{y0,z}\right)s_z \ . \quad (7)$$
$$\frac{2}{c}\frac{ds_z}{dt} = \eta^{xx}h_{0x,z}s_x + \eta^{yy}h_{0y,z}s_y$$

The Green's function in cylindrical polar coordinates can be cumbersome to work with during integration; however, since we are interested in the interior points very near to the optical axis, the Green's function can be expanded using a Taylor-Maclaurin series expansion about the optical axis where the test particle will be placed,

$$G(x, x') = \frac{1}{4\pi}\frac{1}{R_0}\sum_{n=0}^{\infty}\left(\frac{r}{R_0}\right)^n P_n\left(\frac{r'}{R_0}\cos(\theta - \theta')\right). \quad (8)$$

Here $R_0 = R(0,0,z) = \sqrt{r'^2 + (z-z')^2}$ and $P_n(x)$ are the Legendre polynomials. The primed quantities are those of the source of radiation and the unprimed quantities are for the observation points. Equation (8) simplifies calculation by allowing for the trigonometric function to be made separable from the radial $r$ and longitudinal $z$ coordinates.

Even with the simplifications resulting from the use of a steady beam and the expansion of the Green's function in terms of the Legendre polynomials, the integral in Eq. (4) is challenging. Up to this point, we are still within the framework of the paraxial and General Relativistic approximations. Our first approximation is to take the doughnut-shaped intensity profile of the Laguerre-Gaussian beams as a thin cylindrical shell of radius $r_0$ and length $L$ [see Fig. 2],

$$|\psi|^2 = A\frac{\delta(r'-r_0)}{2\pi r_0}\Pi_L(z'). \quad (9)$$

Here $A = \pi r_0^2$ is taken to be the transverse area of the beam, $\delta(r - r_0)$ is the Dirac delta function, and $\Pi_L$ is a top-hat function of length $L$ and centered on $z = 0$ [14]. This type of approximation is adopted from calculations made in Refs [7—9] and is justifiable since the "center-of-mass" along the radial coordinate is located approximately at peak in intensity (i.e., along the ring). The radial position $r_0$ of the peak in the intensity profile $|\psi|^2$, which is determined by the condition $\partial |\psi|^2 / \partial r = 0$, is at $r_0 = w(z)\sqrt{|\ell|/2}$. Because the integral in Eq. (4) is performed with the Dirac delta function of Eq. (9), we can simplify the derivative term appearing in Eq. (3); this derivative is $\partial |\psi|^2 / \partial r = 2(|\ell|/r - 2r/w^2)|\psi|^2$ and upon substituting $r_0 = w(z)\sqrt{|\ell|/2}$ for $r$ into this expression we find $\partial |\psi|^2 / \partial r = 0$. Lastly, we neglect the effects of diffraction by setting $z$-dependent beam parameters to their value at $z = 0$, viz., $w(0) = w_0$ and $R(0) \to \infty$ for the entire track of the beam. The $z$-dependent coefficients in Eq. (7) can then be found by performing the integrals in Eq. (4) with Eqs. (8) and (9), taking the appropriate derivatives in Eq (7), and evaluating the results at $r = 0$,

$$\Gamma_{x0}^{y}(z) = -\Gamma_{y0}^{x}(z) = -\ell \rho_L \frac{G}{\pi c^4} \frac{\lambda}{r_0^2} \left[ \frac{L/2 - z}{\sqrt{r_0^2 + (L/2 - z)^2}} + \frac{L/2 + z}{\sqrt{r_0^2 + (L/2 + z)^2}} \right]. \qquad (10)$$

All other connection coefficients vanish. Here $\rho_L$ is the linear radiation density and is related to the volume density by $\rho_L = \rho_V \pi r_0^2$.

Using the quantities found in Eq. (10), the time rate of change of the spin vector is found to satisfy the cross product $ds/dt = \dot{\Omega} \times s$, where $\dot{\Omega} = (0, 0, \Omega_z^{OV})$ is the precession frequency. The spinning test particle placed at the center of the OV will have a rate of precession given by $\Omega_z^{OV} = c\Gamma_{x0}^{y}(z)$. Our first observation is that the precession frequency in Eq. (10) is "quantized" with the orbital angular momentum mode number of the OV and depends on its sign. Second, for an infinitely long beam track $L \to \infty$ the precession frequency $\Omega_z^{\text{OV}}$ is finite. This is expected since the strength of the "influence" decreases from a finite source. For example, in Ref [9], we find that the strength of the influence on a particle placed at a position $z$ is $\propto a^3 / \left[ \sqrt{2a^2 + 4z^2} \left( a^2 + 4z^2 \right) \right]$, which falls off as $1/z^3$. In the present calculations, for a finite cylinder a Taylor series expansion of the bracketed expression in Eq. 10 at infinity shows that the influence has a $1/z^3$ dependence for large $z$. This dependence is indicative of the frame dragging effect. Positioning the test particle at $z=0$, the strength of the influence is $f = L/\sqrt{4r_0^2 + L^2}$. Letting the beam track to be equal to twice the Rayleigh range (diffraction length) $L = 2z_0$ and allowing the transverse beam size to be equal to $\lambda$, the strength reaches ~95% of its maximum value. This justifies neglecting diffraction and allows us to simplify the expression for the rate of precession as $\Omega_z^{\text{OV}} \approx \pi^{-1} \ell G \rho_L \lambda / r_0^2 c^3$.

The rate of precession in the case of the ring laser was found to be $\Omega_z^{\text{ring}} = 8\sqrt{2} G \rho_L / ac^3$, where $a$ is the length of each side of the ring laser. From the previous discussion on OV's, it was shown that the influence is a collective effect from the beam track, while the influence for the ring laser is from the pointing vector being in the plane of the laser. Both configurations benefit from decreasing the radius of their rings. However, optical vortices can be tightly focused to roughly a wavelength so that $\Omega_z^{\text{OV}} \approx \pi^{-1} \ell G \rho_L / \lambda c^3$. Ultimately, the ring laser dimensions are limited by the damage threshold ($\sim 10^{12}$ W/cm$^2$) of the optical material used in its construction: making the ring laser smaller requires making the beam smaller, which results in a greater peak laser intensity.

To get a feel for the magnitude of the frame-dragging effect, an estimate of the peak laser intensity needed for a 1 Hz precession frequency is made for these "steady configurations". By focusing an 800 nm, single cycle pulse down to a wavelength, the required intensity is

$\sim 10^{45}$ W/cm$^2$. Currently, the world's most intense laser pulses are produced by the Hercules laser system at the University of Michigan, which when delivering its 30 fs laser pulses to a focal spot-size of about a micron yields an intensity of $\sim 2 \times 10^{22}$ W/cm$^2$ [17]. These laser parameters, notably the duration and the intensity of the laser pulses, demonstrates the smallness of this effect.

In conclusion, General Relativistic calculations in the weak-field limit have shown that optical vortices can exhibit frame dragging on a spinning test particle. The precession frequency of the particle was found to be quantized with and dependent of the sign of the OV's topological charge. As a result of the azimuthal component of the Poynting vector, OVs do not require optical constraints to circulate the radiation. Compare to the ring laser configuration, their ring size is limited only by the wavelength of light.

REFERENCES

References with Titles

1) Fowles G R and Cassidy G 1990 Analytical Mechanics 6th ed. (Fort Worth: Saunders)

2) Coles, Peter, "Einstein, Eddington, and the 1919 Eclipse," in V.J. Martinez, V. Trimble & M.J. Pons-Borderia, eds., Proceedings of International School on the Historical Development of Modern Cosmology, Valencia 2000 (Astronomical Society of the Pacific, San Francisco, CA, 2001) pp. 21—41.

3) J. G. von Soldner, "On the deflection of a light ray from its rectilinear motion," B. A. J. 161-172 (1804).

4) A. Einstein, "Die Grundlage der allgemeinen Relativitatsheorie," Ann. Phys. (Leipzig) 49, 769 (1916).

5) F. W. Dyson, A, S, Eddington and C. Davidson, "A determination of the deflection of light by the sun's gravitational field, from observation made at the total eclipse of May 29, 1919," Phil. Trans. R. Soc. Lond. A 220, 291-333 (1920).

6) S. S. Doeleman, et. al., "Event-horizon-scale structure in the supermassive black hole candidate at the Galactic Center," Nature 455, 78 (2008).

7) R. C. Tolman, P. Ehrenfest, and B. Podolsky, "On the gravitational field produced by light," Phys. Rev. 37, 602 (1931).

8) M. O. Scully, "General-relativistic treatment of the gravitational coupling between laser beams," Phys. Rev. D 19, 3582 (1979).

9) R. L. Mallet, "Weak gravitational field of the electromagnetic radiation in a ring laser," Phys. Lett. A 269, 214-217 (2000).

10) R. L. Mallett, "The gravitational field of a circulating light beam," Found. Phys. 33, 1307 (2003).

11) K. D. Olum and A. Everett, "Can a circulating light beam produce a time machine," Found. Phys. Lett. 18, 379 (2005).

12) L. Allen, S. M. Barnett, and M. J. Padgett, eds., Optical Angular Momentum (IOP, 2003).

13) D. J. Griffiths, Introduction to Electrodynamics (Prentice Hall, Upper Saddle River, NJ, 1999), 3rd ed.

14) J. D. Jackson, Classical Electrodynamics (Wiley, New York, 1999), 3rd ed.

15) M. P. Hobson, G. P. Efstathiou, and A. N. Lasenby, General Relativity: An Introduction for Physicists (Cambridge University Press, Cambridge, England, 2006).

16) T. Tajima and G. Mourou, "Zettawatt-exawatt lasers and their applications in ultrastrong-field physics," PRST-AB 5, 031301 (2002).

17) V. Yanovsky, et. al., "Ultra-high intensity-300-TW laser at 0.1 Hz repetition rate," Opt. Express 3, 2109 (2008).

Figure Captions

Fig. 1. (color online). Illustration of a massive spinning neutral particle placed at the center of an optical vortex beam. The multicolored "spiral staircase" surface is the phasefront of an optical vortex beam. The intensity profile of the doughnut-shaped beam is encoded in the transparency of this surface. The blue sphere at the center of the image is a spinning test particle. Its spin vector is shown by the red arrow. The Pointing vector (not shown) follows integral curves which spiral in an opposite sense to that in which the phasefront appears to spiral. In this image the spin precession is in a clockwise sense when viewed from above.

Fig. 2. (color online) A thin cylindrical shell representation of a "steady" Laguerre-Gaussian beam of length $L$ and radius $r_0$ centered on the origin. The transverse beam profile is approximated using a Dirac delta function and the length of the cylinder is approximated using a top-hat function. The optical axis of the beam is along the $z$-axis, and the transverse dimensions of the beam are in the $x$-$y$ plane. The $\vec{x}'$ vector labels the coordinates of the source of radiation on the cylinder. The vector $\vec{x}$ gives the coordinates of the observation points and the vector $\vec{R}$ is the displacement vector between the source and observation points.

Figures

Figure 1

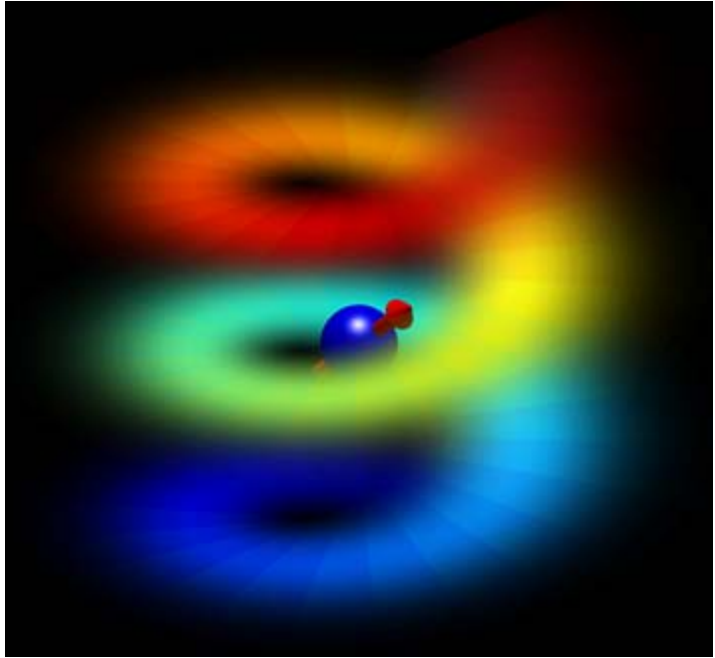

Figure 2

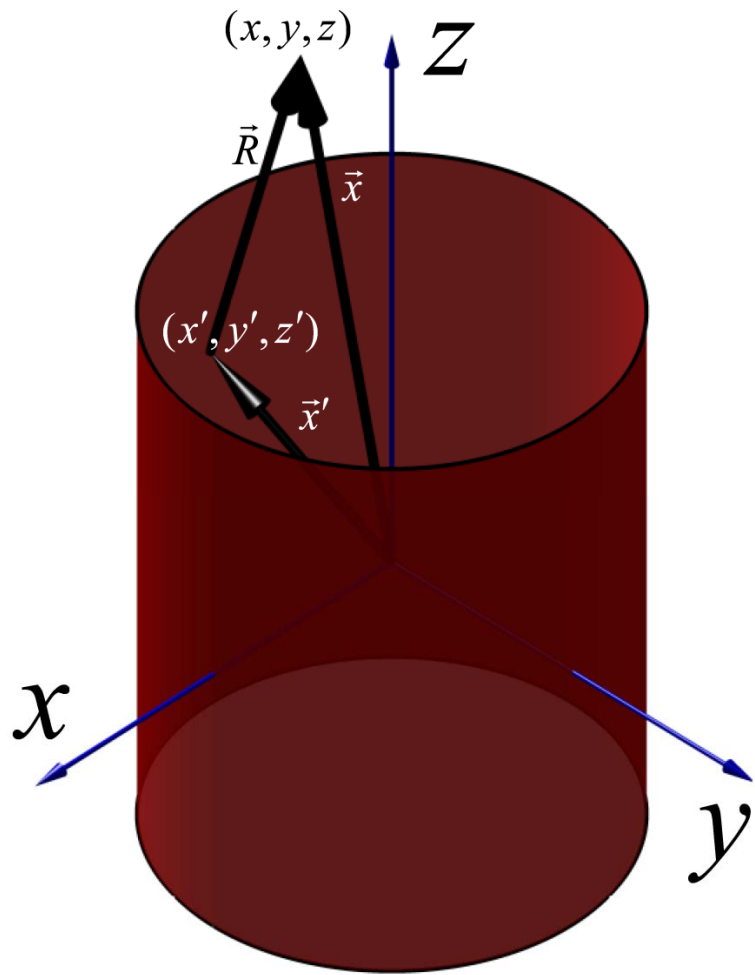